\documentclass[reprint,amsmath,amssymb,aps]{revtex4-1}
\usepackage{graphicx}% Include figure files
\usepackage{bm}% bold math
\usepackage{tcolorbox} % for colour boxes
\usepackage{float} %hopefully forces image positions.
\usepackage{longtable}
\usepackage[caption=false]{subfig}
\usepackage{tabularx}
\usepackage{color}
\usepackage{hyperref}
\usepackage{subfig}
\usepackage[toc,page]{appendix}
\usepackage{soul}   % Packages added by Alberto
\usepackage{tikz}
\usetikzlibrary{shapes.geometric,arrows,automata,positioning}
\usepackage{pgfplots}
\usepackage{xcolor}
\usepgfplotslibrary{groupplots}
\pgfplotsset{width=10cm,compat=1.8}
\usepgfplotslibrary{fillbetween}
\usetikzlibrary{patterns}
\usepackage{tikz-3dplot}
\usepackage{microtype}

\begin{document}
\title{Unified Mechanism of Atrial Fibrillation in a Simple Model}

\date{\today}
             
\author{Max Falkenberg$^{1,2,3}$}
\author{Andrew J. Ford$^{1}$}
\author{Anthony~C.~Li$^{1}$}
\author{Alberto Ciacci$^{1,2,3}$}
\author{Nicholas S. Peters$^{3}$}
\author{Kim Christensen$^{1,2,3}$}
\email[Corresponding author: ]{k.christensen@imperial.ac.uk}
\affiliation{$^1$Blackett Laboratory, Imperial College London, London SW7 2AZ, United Kingdom}
\affiliation{$^2$Centre for Complexity Science, Imperial College London, London SW7 2AZ, United Kingdom}
\affiliation{$^3$ElectroCardioMaths Programme, Imperial Centre for Cardiac Engineering, Imperial College London, London W12 0NN, United Kingdom}

%\begin{abstract}
%\noindent \textbf{Keywords:} atrial fibrillation, arrhythmia, cellular automata, re-entrant circuits, unifying mechanism
%\end{abstract}

\maketitle

%\textbf{Significance Statement: The mechanisms of atrial fibrillation (AF) are poorly understood. Recent clinical work has suggested a potential unifying mechanism, based on drivers forming in the walls of the atria. Our model is the first to qualitatively reproduce the electrical activation patterns observed in the clinical study, and it explains their theoretical origin based on the discrete nature of the atrial fibre network. In addition, the model provides wholly novel insights, predicting that the drivers of AF move deeper into the heart wall as AF becomes more persistent. Drivers further from the inner atrial wall would be harder to destroy with ablation. Hence, this prediction, if clinically confirmed, may explain why persistent AF has lower treatment success rates than paroxysmal (intermittent) AF.}

\textbf{The mechanism of atrial fibrillation (AF) is poorly understood, resulting in disappointing success rates of ablative treatment. Different mechanisms defined largely by different atrial activation patterns have been proposed and, arguably, this dispute has slowed the progress of AF research. Recent clinical evidence suggests a unifying mechanism based on sustained re-entrant circuits in the complex atrial architecture. Here, we present a simple computational model showing spontaneous emergence of AF that strongly supports, and gives a theoretical explanation for, the clinically observed diversity of activation. We show that the difference in surface activation patterns is a direct consequence of the thickness of the discrete network of heart muscle cells through which electrical signals percolate to reach the imaged surface. The model naturally follows the clinical spectrum of AF spanning sinus rhythm, paroxysmal and persistent AF as the decoupling of myocardial cells results in the lattice approaching the percolation threshold. This allows the model to make additional predictions beyond the current clinical understanding, showing that for paroxysmal AF re-entrant circuits emerge near the endocardium, but in persistent AF they emerge deeper in the bulk of the atrial wall where endocardial ablation is less effective. If clinically confirmed, this may explain the lower success rate of ablation in long-lasting persistent AF. %This model is unique in its simplicity and ease for generating hypotheses to be tested clinically. 
%\cite{lee2015,zaman2018,swarup2014}. 
%\cite{calkins2017}.
%In the future, the model can be applied to gain a systematic understanding of the underlying mechanisms of AF and develop and improve clinical treatment strategies.
}

%%%%%%%%%%%%%%%%%%%%% BEGIN FIGURES %%%%%%%%%%%%%%%%%%%%%
%%%%% FIGURE 1 %%%%%
\begin{figure}
\centering
\includegraphics[width=\linewidth]{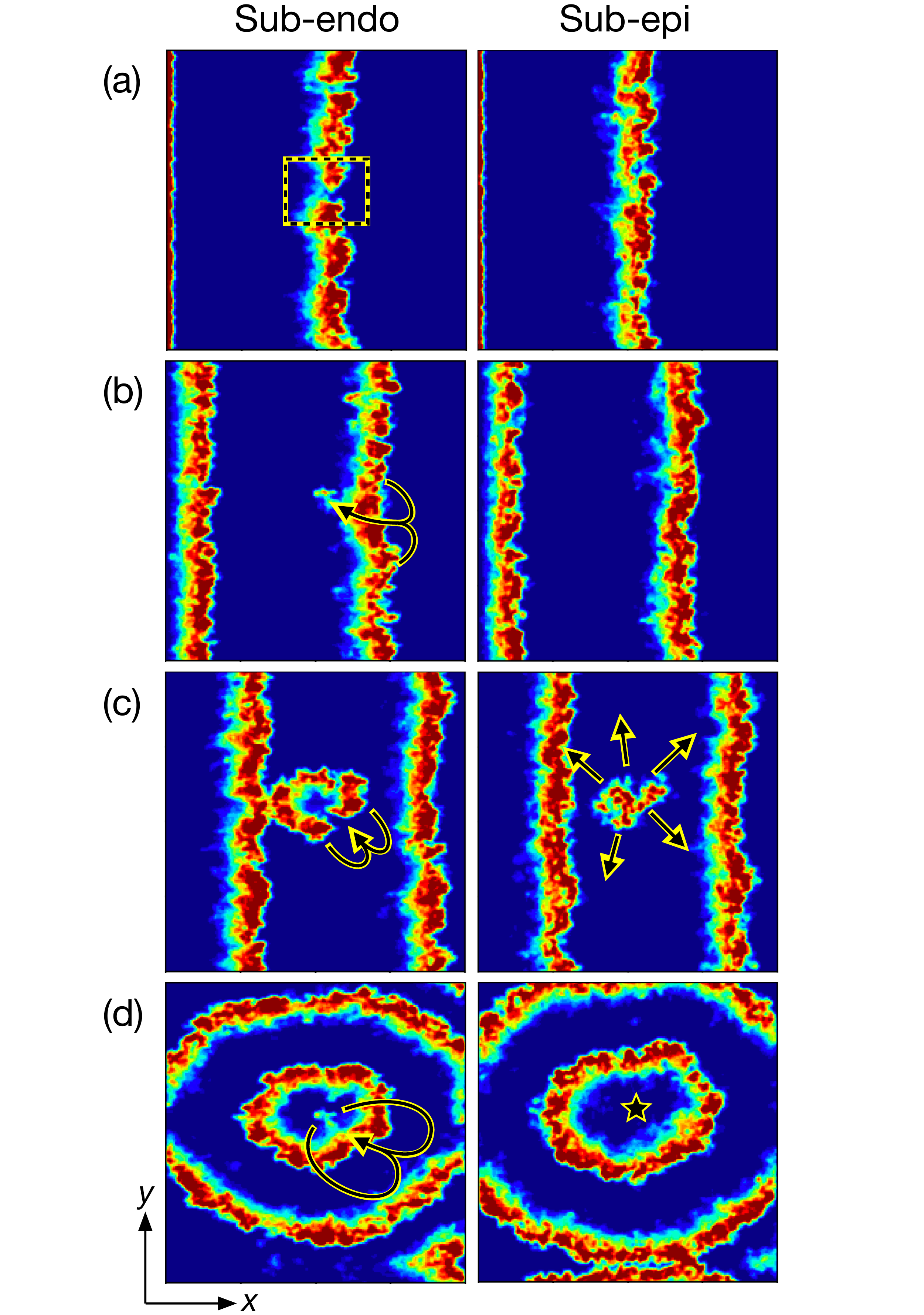}
\caption{The emergence of AF in the inhomogeneous model with simultaneous endocardial ($z=0$) and epicardial ($z=24$) imaging. Gaussian smoothing has been applied for clarity. Red: excited nodes, blue: resting (excitable) nodes, other: refractory (unexcitable) nodes. (a) Planar wavefronts propagating during sinus rhythm. Endo view: Dotted box indicates gap in wavefront formed by conduction block. (b) Endo view: Arrow indicates excitation re-entering gap in wavefront. Epi view: Re-entry is not observed. (c) Emergence of fibrillatory activity. (d) Maintenance of fibrillatory activity. Epi view: Activity emerges on surface as point source located at the star. See videos A in supplementary information.}
\label{fig:3dCMP}
\end{figure}

%%%%% FIGURE 2 %%%%%
\begin{figure*}
\centering
    \subfloat{\label{fig2sublable1}\includegraphics[keepaspectratio,width=9cm]{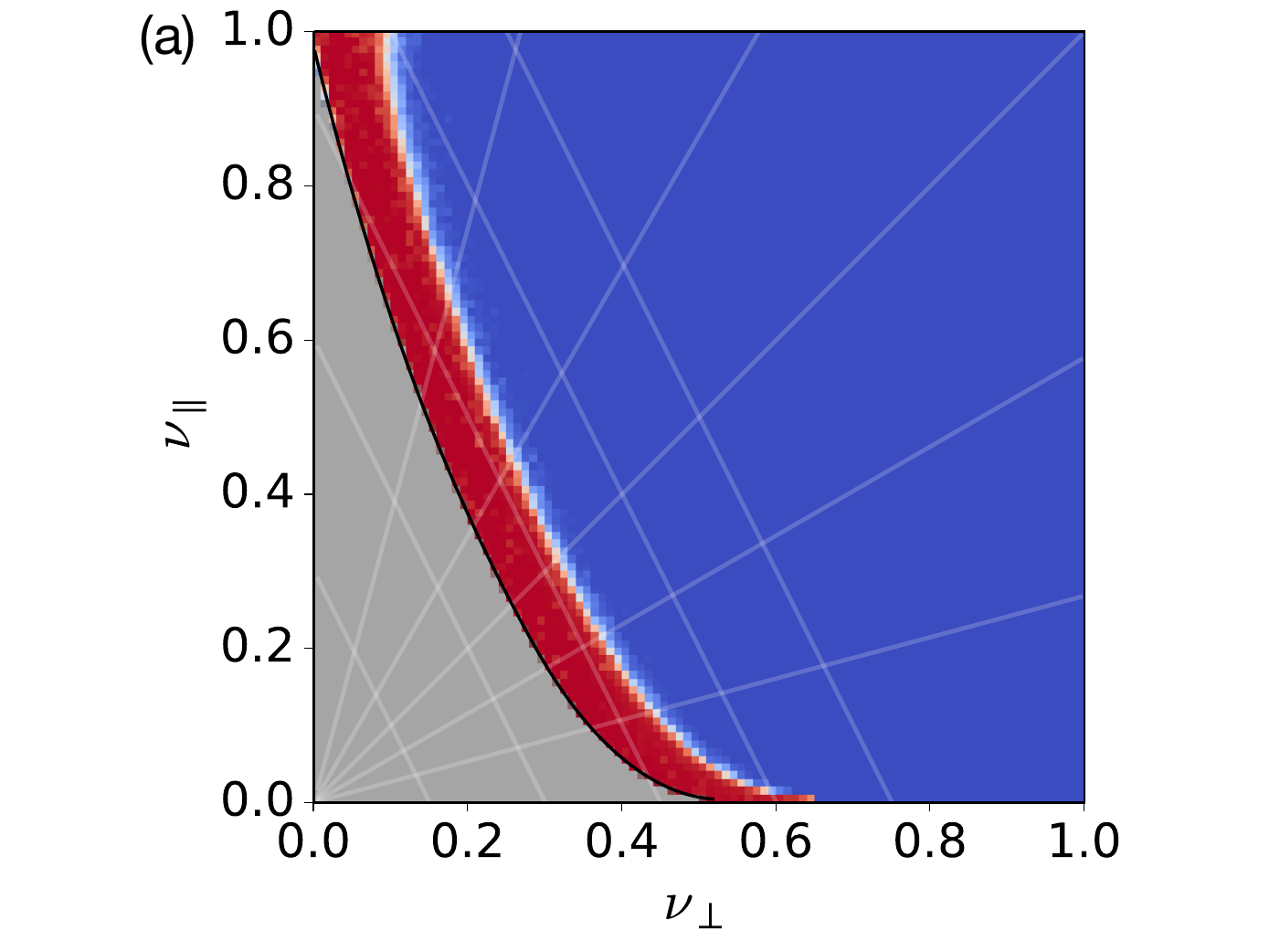}} 
    \subfloat{\label{fig2sublable3}\includegraphics[keepaspectratio,width=9cm]{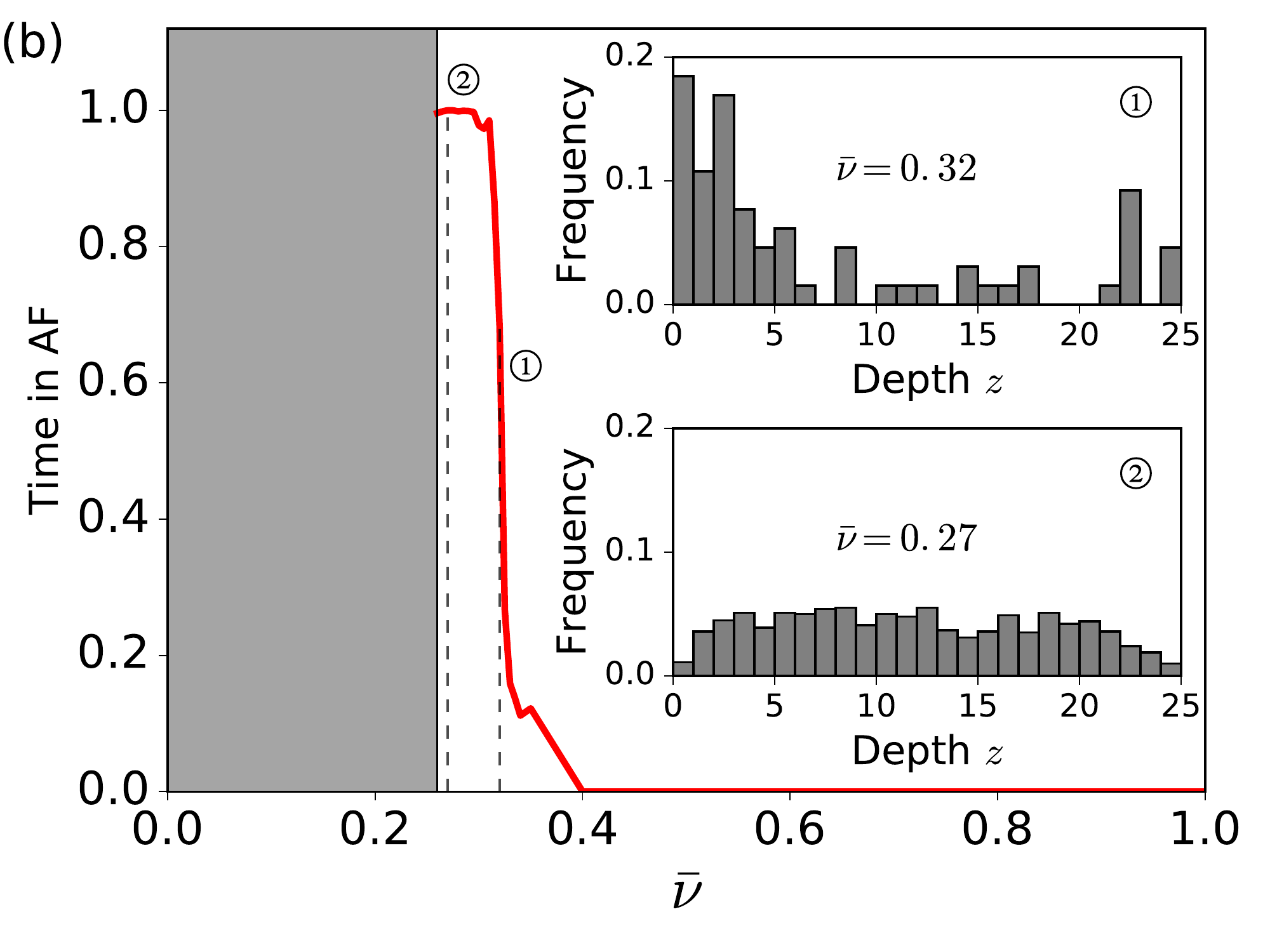}}
\vspace*{-0.4cm}
    \caption{The average time spent in AF for a model with (a) homogeneous coupling parameters, and (b) inhomogeneous coupling parameters with a fixed average coupling, $\bar{\nu}$. Both models show a transition from sinus rhythm at large coupling (0\% AF time) through paroxysmal AF (0\% $<$ AF time $<$ 100\%) to persistent AF (100\% AF time). The grey parameter regions, indicate coupling parameters where there is no spanning network of muscle cells from the sinus node at $x=0$ to $x = L_x$, and are thus irrelevant to clinical AF. (a) Homogeneous coupling parameters. Blue: sinus rhythm, red: persistent AF. The white transition region corresponds to paroxysmal AF. Guidelines of positive (negative) gradient indicate constant $\Delta \theta$ $\left(\bar{\nu}\right)$. (b) Inhomogeneous coupling parameter risk curve (red graph). There is no AF for $\bar{\nu} \gtrsim 0.4$. Decreasing $\bar{\nu}$ associated with decoupling cells, there is a transition from sinus rhythm through paroxysmal AF to persistent AF. Inset: Histograms showing the distribution of re-entrant circuits driving AF for paroxysmal AF (top) and persistent AF (bottom) as a function of depth $z$ from the endocardium. Drivers cluster in the sub-endocardial region ($z=0$) for paroxysmal AF but are uniformly distributed across the bulk for persistent AF.}
    \label{fig:riskcurves}
\end{figure*}

%%%%% FIGURE 3 %%%%%

%%%%%%%%%%%%%%%%%%%%%%%%%%%%%% END FIGURES %%%%%%%%%%%%%%%%%%%%%%%%%

Atrial fibrillation (AF) is the most common cardiac arrhythmia and is projected to affect 1\% of the world's population by 2030 \cite{patel2018}. The key disagreements in the mechanistic understanding of AF concern whether the sustaining mechanism of AF is due to localised drivers of fibrillatory activity or the random motion and interaction of multiple wavelets across the atria \cite{nattel2002,schotten2016,federov2018,nattel2017B,Mann2018}. As a result, treatment by ablation still has disappointing success rates \cite{verma2015,macle2015,narayan2017}. There is growing evidence in favour of local drivers as a sustaining mechanism of AF \cite{lee2015,narayan2013,haissaguerre2014,narayan2017,hansen2015,hansen2016,csepe2017,federov2018}, with a recent meta-analysis \cite{baykaner2018} demonstrating that ablating local drivers has an additional benefit over pulmonary vein isolation, the most common ablation strategy \cite{haissaguerre1998}. 
Additionally, it has been shown that persistent AF is only terminated when local drivers are targeted \cite{zaman2018}. However, there is debate about the mechanistic origin of local drivers:
Some studies have primarily identified the local drivers as rotors using phase analysis \cite{narayan2013,haissaguerre2014}, whereas another study has identified the drivers of AF as focal or breakthrough points \cite{lee2015}. 

%\begin{figure}
%\centering
%\includegraphics[width=\linewidth]{hansen_combi}
%    \caption{\label{fig3}(a) The distribution of activation patterns observed from simultaneous mapping of the endocardium and the epicardium. LRA\# refer to the patient number in the study \cite{hansen2015}. The figure shows that re-entry activity is predominantly observed on the endocardial surface, whereas focal sources are only seen on the epicardium. (b) An activation map from the clinical study \cite{hansen2015} showing re-entry activity from sub-endo imaging and a focal source on the sub-epi side with four marked reference points. (c) The equivalent as (b) for our model. Gaussian smoothing has been applied to images for clarity. Red: excited nodes, blue: resting (excitable) nodes, other: refractory nodes. (a) and (b) have been adapted with permission from Hansen \textit{et al.} \cite{hansen2015}.}
%    \label{fig:hansen}
%\end{figure}

\begin{figure}
\centering
\includegraphics[width=\linewidth]{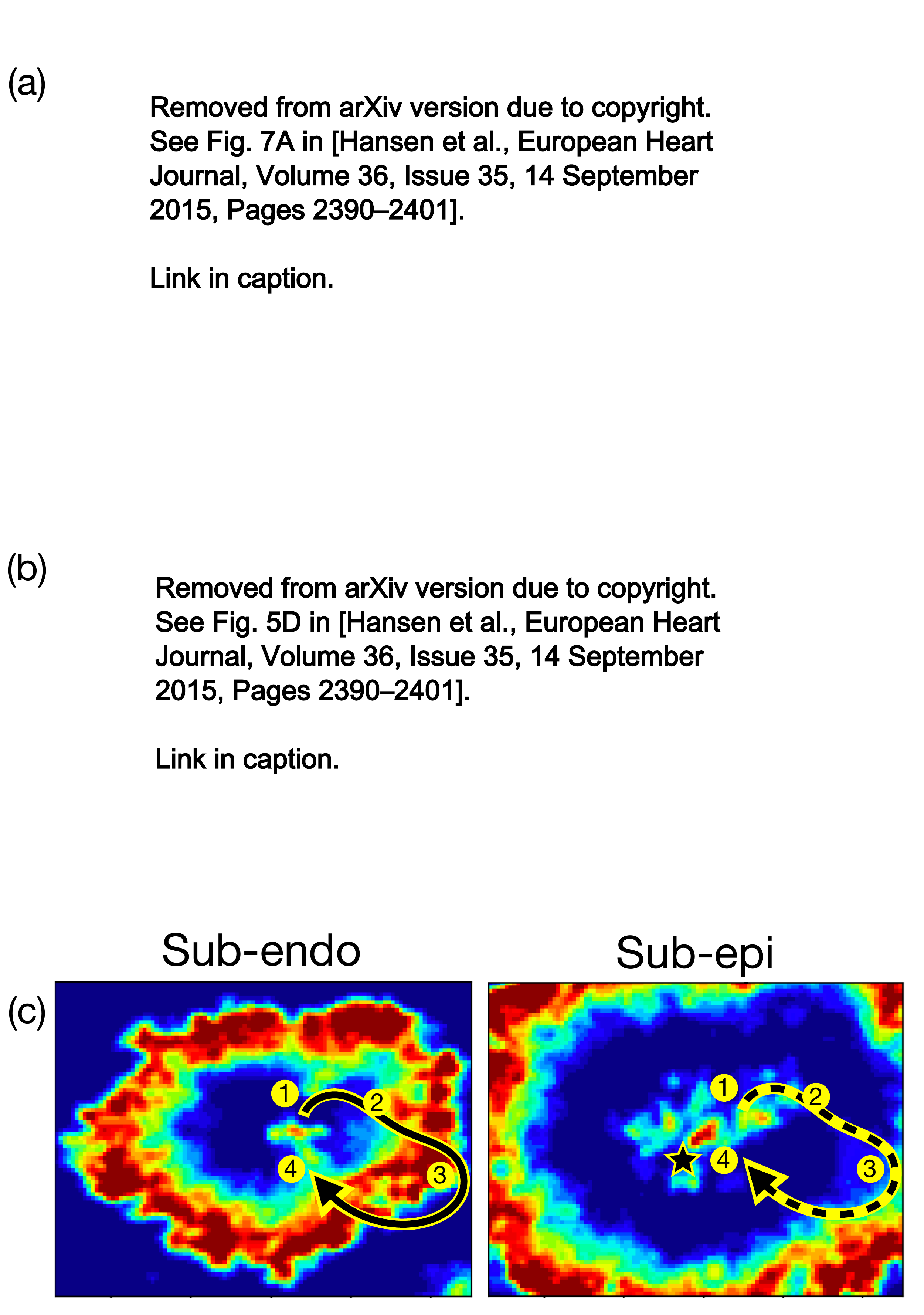}
    \caption{\label{fig3}(a) The distribution of activation patterns observed from simultaneous mapping of the endocardium and the epicardium. LRA\# refer to the patient number in the study \cite{hansen2015}. The figure shows that re-entry activity is predominantly observed on the endocardial surface, whereas focal sources are only seen on the epicardium. (b) An activation map from the clinical study \cite{hansen2015} showing re-entry activity from sub-endo imaging and a focal source on the sub-epi side with four marked reference points. (c) The equivalent as (b) for our model. Gaussian smoothing has been applied to images for clarity. Red: excited nodes, blue: resting (excitable) nodes, other: refractory nodes. \textbf{(a) and (b) have been removed from arXiv version. See Figs.~7A and 5D respectively in \cite{hansen2015}. For links to figures click \href{https://academic.oup.com/view-large/figure/86342521/ehv23307.tiff}{HERE} and \href{https://academic.oup.com/view-large/figure/86342516/ehv23305.tiff}{HERE}.}}
    \label{fig:hansen}
\end{figure}

Recently, a unifying mechanism has been proposed, suggesting that all local drivers may be explained by the presence of small re-entrant circuits \cite{hansen2015,nattel2017,calkins2017}. These re-entrant circuits have a spatial extent of approximately 15mm \cite{hansen2015}.
Using high-resolution simultaneous optical mapping of the endocardium (inner layer of heart wall) and epicardium (outer layer of heart wall) in explanted, diseased human hearts, the study shows that stable transmural re-entrant circuits may project differently onto the endocardium and the epicardium. Projections may appear as rotational activity, or as stable or unstable breakthrough points. Hence, the apparently incompatible 2d projections onto the endocardium and epicardium are consistent with a single underlying mechanism of localised 3d re-entrant circuits in the transmural region \cite{hansen2015,hansen2016,csepe2017}.

In this paper, we demonstrate that these seemingly incompatible surface activation patterns are two sides of the same coin. We introduce a model in which activation wavefronts propagate in a 3d medium mimicking the complex discrete fibre structure of the atria. AF emerges spontaneously in the model through the formation of spatially stable but temporarily intermittent re-entrant circuits. The model predicts that these re-entrant circuits should have a spatial extent just above a lower bound of 12.5mm which is directly compatible with Ref.~\cite{hansen2015}.

A strength of this model is that despite being very simple,
it successfully demonstrates the range of activation patterns observed on the endocardium and the epicardium during AF, closely matching the key clinical observations in Ref.~\cite{hansen2015} as well as several other observations in the recent AF literature \cite{lee2015,zaman2018,swarup2014}. The observed activation patterns fundamentally change as a function of the depth from the imaged surface to the driving re-entrant circuit. If this depth is large, we observe focal or breakthrough activity. If this depth is small, we observe re-entrant or rotational activity.
Hence, our model in combination with the clinical results in Ref.~\cite{hansen2015} provides strong evidence for this elegant unifying mechanism of AF. Furthermore, the model predicts that the drivers of paroxysmal AF are predominantly located near the endocardium whereas the drivers of persistent AF are uniformly distributed within the bulk of the atrial wall. Consistent with clinical experience, the former is more easily ablated than the latter. 
The simplicity of our model, as opposed to highly complex models of AF \cite{clayton2011}, makes it
ideal for testing mechanistic hypotheses, proposed treatment and preventive strategies, and targeting and delivering of ablation. Therefore, in the future, the model may be used as a test bed for unlocking new promising intervention procedures.

\textbf{Model.} %The membrane potential of heart muscle cells shows that they can be in one of three states: active, where the cell depolarises, resting, where the cell can be depolarised by a neighbouring active cell, and refractory, where for a short time after excitation the cell cannot be re-excited. 
Electrical signals in the heart are mediated by discrete heart cells arranged in long, intertwined fibres. This motivates the construction of a 3d cellular automata model of AF where each node acts as a single cell (or block of cells), and nodes are connected stochastically to their neighbours to mimic the branching structure of discrete heart muscle fibres, generalising a 2d model \cite{christensen2015}. 

Discrete models of cardiac tissue have gained popularity recently \cite{alonso2013,christensen2015,vigmond2016,alonso2016,lin2017}. In these models,  the onset of re-entry has been associated with the approach from above of the bond occupation probability to the percolation threshold. Some models have been extended to 3d to highlight the importance of the thickness of the atrial wall to the probability of re-entry \cite{alonso2016}. Additionally, leading clinicians have highlighted the importance of bond percolation to the emergence of AF, noting that the appearance of complex fractionated atrial electrograms can be explained by percolation \cite{vigmond2016}. 

In our model, we consider a simplified $L_{x} \times L_{y} \times L_{z}$ pipe topology of the atria with open boundaries in $x$ and $z$ and periodic boundaries in $y$. Nodes are connected longitudinally to their neighbours in the $x$ direction with frequency $\nu_{\parallel}$ and transversely in the $y$ and $z$ directions with frequency $\nu_{\perp}$. 
%In the real atria, coupling is much stronger longitudinally than transversally, mimicking the strong electrical coupling along muscle fibres, and weaker coupling across fibres.

Each node can take one of three states: resting, where the node can be excited by an active neighbour, excited, or refractory, where for $\tau$ time steps after excitation a node cannot be re-excited. We define the sinus node (pacemaker) as nodes at the boundary, $x=0$, which excites every $T$ time steps. Model parameters $L_{x} = L_{y} = 200$, $L_{z} = 25$, $T = 220$ and $\tau = 50$ are informed by clinically observed values, see supplementary information. We define a small fraction of nodes, $\delta$, as representing locations where cells are susceptible to conduction block. With small probability $\epsilon$, these nodes fail to activate when their neighbours excite. When the sinus node excites in normal conduction, wavefronts are initiated at $x = 0$ and propagate smoothly in the $+x$ direction and terminate at $x=L_{x}$. However, re-entrant circuits can form when nodes are sufficiently decoupled, through fibrosis \cite{spach2006} or otherwise, such that the shortest closed loop from a node back to itself is partially isolated from the remaining tissue, and the path length, in units of propagation time, exceeds the refractory period $\tau$. This gives a minimum spatial extent of $\tau/2$ which, when converted back into clinical units, corresponds to 12.5mm.
%The re-entrant circuit is wrapped around a line of conduction block.
The formation of these re-entrant circuits, wrapped around a line of conduction block, initiate and maintain (drive) fibrillatory activity where the electrical activity is disorganised within the atria \cite{christensen2015,McGillivray2018}, see Fig.~\ref{fig:3dCMP}.

A key observation in Ref.~\cite{hansen2015} is that re-entry activity is typically observed on the endocardial surface, whereas focal or breakthrough points are often found on the epicardial surface. Additionally, there is much stronger longitudinal coupling on the endocardial surface than on the epicardial surface. To test whether fibre orientation can account for the distribution of activation patterns observed clinically, we consider a homogeneous and inhomogeneous model. In the homogeneous model all nodes are connected to their neighbours with the same combination of frequencies $\nu_{\parallel}$ and $\nu_{\perp}$. In the inhomogeneous model, we allow the variation in fibre direction to change with depth. Here we fix the average decoupling of nodes, $\bar{\nu} = (2\nu_{\parallel} + 4\nu_{\perp})/6$ and vary linearly the average fibre angle, $\Delta \theta = \tan^{-1}(\nu_{\perp}/\nu_{\parallel})$, in each layer from $\Delta \theta_\mathrm{ENDO} = 24^{\circ}$ at the endocardium ($z=0$) to $\Delta \theta_\mathrm{EPI} = 42^{\circ}$ at the epicardium ($z=24$), informed by \cite{csepe2017}.
%however, as a first approximation we neglect the global average in determining the fibre orientation.

\textbf{Results.} We have studied the emergence of AF in the model as a function of $\nu_{\parallel}$ and $\nu_{\perp}$ for the homogeneous model and as a function of $\bar{\nu}$ for the inhomogeneous model.

The phase space for the risk of entering AF for homogeneous model is shown in Fig.~\ref{fig2sublable1} and for the inhomogeneous model in Fig.~\ref{fig2sublable3}. For large values of the coupling parameters (no fibrosis) the model exhibits sinus rhythm indefinitely. As the coupling reduces, due to fibrosis or otherwise, a transition takes place where a small number of re-entrant circuits can form. Here we observe paroxysmal AF with intermittent episodes of irregular activity. As the coupling is even further reduced, the model enters persistent AF where, once AF has been initiated, the model will never return to sinus rhythm without external intervention.
These results are consistent with recent evidence showing that local drivers anchor at or near fibrotic lesions \cite{haissaguerre2014}, with fibrosis increasing the number of re-entrant regions in the atria, and increasing the duration spent in AF \cite{cochet2018}.

%Re-entry activity is observed from surface imaging when the driver is close to the surface being imaged. Conversely, focal activity is observed if the driver is far from the surface being imaged. 
In the model, we can quantify where re-entrant circuits emerge in the substrate and observe the associated activation patterns on the endocardial and epicardial surfaces. For the homogeneous model variant there is no distinction between the endocardium and the epicardium, and we find that in paroxysmal AF, the majority of drivers form equivalently on either surface and not in the bulk, whereas for persistent AF, the majority of drivers form in the bulk and not on the surfaces, see Fig.~11 in supplementary information. 

However, the inhomogeneous model breaks the symmetry between the epicardium and the endocardium, consistent with clinical observations \cite{csepe2017}. Here we find that in paroxysmal AF, drivers form preferentially near the endocardial surface, with very few drivers forming on the epicardial surface and almost none in the bulk of the atria, see Fig.~\ref{fig2sublable3}, inset 1. However, in persistent AF, drivers are uniformly distributed throughout the atrial wall, see Fig.~\ref{fig2sublable3}, inset 2. Hence, as AF becomes more persistent, the average position of drivers moves away from the endocardium and into the bulk of the atrial wall. The surface activity driven from the atrial bulk will typically appear as a breakthrough point, as opposed to re-entry activity observed when drivers are close to the imaged surface. Indeed, in our model we find that ablating the active re-entrant circuits in a tissue terminates AF, consistent with \cite{hansen2015}, see videos F\&G in supplementary information. However, if the ablation does not penetrate far enough into the tissue, fibrillatory activity will continue, see Fig.~14 in supplementary information.

In the 3d model, re-entrant circuits can form anywhere in the bulk of the atrial wall. If we visualise the activation patterns in the inhomogeneous model on the surfaces at $z=0$ (endocardium) and $z=24$ (epicardium), we observe a diversity of activation patterns consistent with the clinical study \cite{hansen2015} which is summarised in Fig.~\ref{fig3}a. Figure~\ref{fig3}b shows a specific example activation map from the clinical study. Figure~\ref{fig3}c shows the equivalent from our model. In Fig.~\ref{fig3}, the left (right) panel shows the activity when viewed from the endocardium (epicardium). For the activation maps in (b) and (c), electrical activity spreads across the tissue from point (1) through point (2), but is blocked from reaching the isolated fibre at point (4). The activity loops around the isolated fibre and re-enters the fibre from the far end at point (3). This back-propagating excitation now passes through point (4) before re-exciting point (1) which sustains the re-entrant circuit. Viewing the same region from the epicardium does not show the same re-entry activity. Instead, the excitation emerges as a point source indicated by a star. In incomplete re-entry, only part of the cycle shown in the left panel is visible from endocardial imaging. For unstable breakthrough, the point at which excitations emerge in the epicardial view vary from beat to beat. For example
activation maps and videos, see supplementary information. 

\textbf{Discussion.} These model observations are directly compatible with recent clinical findings that identified rotational activity from endocardial mapping in patients with paroxysmal AF \cite{swarup2014,narayan2013}, and another clinical observation identifying focal activity in
patients with persistent AF from epicardial mapping \cite{lee2015}. Moreover, our model offers a natural explanation why ablation is more successful for paroxysmal than persistent AF. Clinical studies have shown that the radius of ablation lesions become smaller as they penetrate further into the tissue from the endocardial surface, and that ablation struggles to penetrate more than 2mm into the wall of the atria \cite{Kumar2015} which can be up to 7mm thick \cite{hansen2015}. Hence, because the average driver position moves deeper into the atrial wall as AF becomes more persistent, ablation lesions need to be more accurately positioned and they must penetrate further into the tissue. This restricts the efficacy of AF ablation. 

The model may also explain the prevalence of paroxysmal AF in the pulmonary veins (PVs) \cite{calkins2017,haissaguerre1998}. Sleeves of cardiac tissue extend into the PVs from the atria. The sleeves get thinner, further into the PVs and show many abrupt changes in fibre orientation \cite{Ho1999,weiss2002,weerasooriya2003}. These physiological observations all reduce local cell to cell coupling which the model has shown is the key requirement for forming the re-entrant circuits that initiate and sustain AF.

% Finally, there has been success when ablating rotor cores when using phase mapping \cite{narayan2013,haissaguerre2014}. A very recent study suggests that phase mapping preferentially identifies lines of conduction blocks rather than rotor activity \cite{podziemski2018}. Our model suggests that the success of rotor ablation is because this strategy is, in effect, targeting the lines of conduction block that re-entrant circuits wrap around.

\textbf{Conclusions.}  
The overarching aim was to create the simplest possible bio-inspired model that would recreate the variety of clinically observed activation patterns. 
The model that integrates a coarse grained dynamics of an electrical signal percolating through a complex discrete fibre network mimicking the real 3d structure of the human atria is very simple in its formulation. Nevertheless, we show that a variety of known key characteristics of clinical AF emerge naturally without much of the details thought to be essential in modern AF modelling. In addition to explaining the diversity of activation patterns, the model is consistent with the typical evolution of AF from sinus rhythm through paroxysmal AF to persistent AF. 

Combined with the clinical observations in Ref.~\cite{hansen2015}, our model gives substantial evidence for the proposed unifying mechanism of AF, suggesting a potential resolution to long-standing debates.
Complementing this finding, we predict that the inhomogeneity in the fibre orientation causes
the average depth of drivers to move from the endocardium and into the bulk of the atrial wall as AF becomes more persistent. We hypothesise that this may explain why ablation is less successful for persistent AF than for paroxysmal AF. This insight is the first of its kind and may have key clinical implications. 
% In the future, the model could be used to test, in a systematic way, a variety of proposed intervention strategies and thereby help to reveal new promising clinical treatment strategies.

In the future, the model will allow us to test the hypothesis that adjusting a small number of parameters underpins the mechanistic origin of AF. Additionally, coupled with newly available high resolution atrial fibre maps \cite{Pashakhanloo2016}, the model will enable rigorous testing and development of treatment and prevention strategies.

\textbf{Author Contributions.} KC conceived the model. MFM proposed the connection to \cite{hansen2015} and adapted the model for variable fibre orientation. AJF \& ACL implemented the model and ran simulations, supervised by MFM. AC developed the theoretical aspects of the model. NSP provided the medical expertise for the project. KC supervised all aspects of project. All authors contributed to the writing of the manuscript.

\textbf{Acknowledgements.} MFM thanks Timotej Kapus and \'Eamonn Murray for computational support. MFM and AC gratefully acknowledge support from the EPSRC. NSP acknowledges funding from the British Heart Foundation (RG/16/3/32175 and Centre of Research Excellence), and the National
Institute for Health Research (UK) Biomedical Research Centre. KC and NSP acknowledge funding from the Rosetrees Trust.

%\clearpage
\bibliography{ref.bib}

\end{document}